\documentclass[journal=apchd5,manuscript=article,layout=twocolumn]{achemso}

\usepackage[utf8]{inputenc}
\usepackage[english]{babel}
\usepackage{microtype}
 
\usepackage{bm}
\usepackage{xcolor}
\usepackage{graphicx}
\usepackage{float}
\usepackage{xspace}
\usepackage{amsmath}
\usepackage{amssymb}
\usepackage{amsfonts}

\def\ii{{\rm i}}  \def\ee{{\rm e}}  
\def\epsa{\epsilon_{\rm a}}  \def\epsb{\epsilon_{\rm b}}
\def\EF{{E_{\rm F}}}    
\def\rb{{\bf r}}  \def\pb{{\bf p}}  \def\Eb{{\bf E}}

\title{Polaritons in two-dimensional parabolic waveguides}

\author{T.~P. Rasmussen}
\affiliation{Center for Nano Optics, University of Southern Denmark, Campusvej 55, DK-5230~Odense~M, Denmark}
\alsoaffiliation{Department of Physics, Chemistry, and Pharmacy, University of Southern Denmark, Campusvej 55, DK-5230~Odense~M, Denmark}

\author{P.~A.~D.~Gon\c{c}alves}
\affiliation{Center for Nano Optics, University of Southern Denmark, Campusvej 55, DK-5230~Odense~M, Denmark}
\alsoaffiliation{ICFO --- Institut de Ci{\`e}ncies Fot{\`o}niques, The Barcelona Institute of Science and Technology, 08860 Castelldefels (Barcelona), Spain}

\author{Sanshui~Xiao}
\affiliation{Department of Photonics Engineering, Technical University of Denmark, DK-2800 Kongens Lyngby, Denmark}
\alsoaffiliation{Center for Nanostructured Graphene, Technical University of Denmark, DK-2800 Kongens Lyngby, Denmark}

\author{Sebastian~Hofferberth}
\affiliation{Department of Physics, Chemistry, and Pharmacy, University of Southern Denmark, Campusvej 55, DK-5230~Odense~M, Denmark}
\alsoaffiliation{Institute of Applied Physics, University of Bonn, Wegelerstra{\ss}e 8, D-53115 Bonn, Germany}

\author{N.~Asger~Mortensen}
\affiliation{Center for Nano Optics, University of Southern Denmark, Campusvej 55, DK-5230~Odense~M, Denmark}
\alsoaffiliation{Danish Institute for Advanced Study, University of Southern Denmark, Campusvej 55, DK-5230~Odense~M, Denmark}
\alsoaffiliation{Center for Nanostructured Graphene, Technical University of Denmark, DK-2800 Kongens Lyngby, Denmark}

\author{Joel D. Cox}
\email{cox@mci.sdu.dk}
\affiliation{Center for Nano Optics, University of Southern Denmark, Campusvej 55, DK-5230~Odense~M, Denmark}
\alsoaffiliation{Danish Institute for Advanced Study, University of Southern Denmark, Campusvej 55, DK-5230~Odense~M, Denmark}

\begin{document}

\begin{abstract}
{\bf
The suite of highly confined polaritons supported by two-dimensional (2D) materials constitutes a versatile platform for nano-optics, offering the means to channel light on deep-subwavelength scales. Graphene, in particular, has attracted considerable interest due to its ability to support long-lived plasmons that can be actively tuned via electrical gating. While the excellent optoelectronic properties of graphene are widely exploited in plasmonics, its mechanical flexibility remains relatively underexplored in the same context. Here, we present a semi-analytical formalism to describe plasmons and other polaritons supported in waveguides formed by bending a 2D material into a parabolic shape. Specifically, for graphene parabolas, our theory reveals that the already large field confinement associated with graphene plasmons can be substantially increased by bending an otherwise flat graphene sheet into a parabola shape, thereby forming a plasmonic waveguide without introducing potentially lossy edge terminations via patterning. Further, we show that the high field confinement associated with such channel polaritons in 2D parabolic waveguides can enhance the spontaneous emission rate of a quantum emitter near the parabola vertex. Our findings apply generally to 2D polaritons in atomically thin materials deposited onto grooves or wedges prepared on a substrate or freely suspended in a quasi-parabolic (catenary) shape. We envision that both the optoelectronic and mechanical flexibility of 2D materials can be harnessed in tandem to produce 2D channel polaritons with versatile properties that can be applied to a wide range of nano-optics functionalities, including subwavelength polaritonic circuitry and bright single-photon sources.
}
\end{abstract}

\hfill



\section{Introduction}

Nanoscopic control of light is of crucial importance both in next-generation photonic technologies and in fundamental explorations at the boundary of classical and quantum physics~\cite{atwater2007promise}. In this context, plasmonics~\cite{maradudin2014modern} harnesses surface plasmons---collective oscillations in the free-electron plasma---at interfacing metallic and dielectric media to concentrate light on length scales far below its associated wavelength in free space~\cite{Gramotnev:2010,Gramotnev:2014}. Research on plasmonics has traditionally relied on the engineering of noble metal nanostructures to tailor plasmon resonances, which can focus light into nanometric volumes for diverse applications including solar energy harvesting~\cite{Atwater:2010}, biosensing~\cite{Brolo:2012}, photocatalysis~\cite{Brongersma:2015}, and plasmonic colors~\cite{Kristensen:2017}. Surface plasmon polaritons (SPPs) have also been explored for their ability to transport electromagnetic energy on subwavelength dimensions~\cite{barnes2003surface}, thus circumventing the limitations of conventional optical waveguides in dielectric-based information and communications technologies~\cite{Ebbesen:2008,fang2015nanoplasmonic}.

Plasmon-based waveguiding has been explored in metallic nanostructures such as wedges and grooves, which effectively create a one-dimensional channel for SPPs---termed channel plasmon polaritons (CPPs) in those structures---that confine light in both lateral and vertical directions~\cite{smith2015gap}. Early research on CPPs dates back to the 1970s, in a theoretical study of the electrostatic modes sustained by a sharp dielectric wedge by Dobrzynski and Maradudin~\cite{dobrzynski1972electrostatic}; their work triggered subsequent explorations of more realistic channel waveguides with rounded apexes, such as parabolic~\cite{eguiluz1976electrostatic,boardman1981retarded,garcia1985excitation} and hyperbolic~\cite{davis1976electostatic,boardman1985electrostatic} channel configurations. In the past two decades, improvements in characterization and nanofabrication techniques have led to numerous experimental investigations of CPPs~\cite{smith2015gap,Bozhevolnyi:2006,moreno2006channel,Sondergaard:2012,Raza:2014,Bermudez-Urena:2015,Bermudez-Urena:2017}.

While noble metals (e.g., Au and Ag) are conventionally used in modern plasmonic waveguiding architectures~\cite{maier2007plasmonics}, their surface plasmon resonances are difficult to tune actively and are limited by relatively large intrinsic Ohmic losses~\cite{Khurgin:2015,Boriskina:2017}. These limitations are uniquely addressed in graphene, a one-atom-thick carbon layer which supports relatively long-lived, highly-confined, and gate-tunable plasmonic excitations in the terahertz to mid-infrared spectral ranges~\cite{GoncalvesPeres:2016,GarciadeAbajo:2014,Xiao:2016,Goncalves:2020a,Goncalves_SpringerTheses}. Graphene plasmons offer tantalizing prospects for nanophotonic functionalities of interest, including the guiding of electromagnetic radiation at  extremely subwavelength scales and strong light--matter interactions for nonlinear and quantum optics~\cite{Koppens:2011,Cox:2019,Goncalves:2020a,Reserbat-Plantey:2021}. In this context, nanostructured graphene in various morphologies has been explored~\cite{GoncalvesPeres:2016,Yu:2017,Goncalves_SpringerTheses}, both theoretically and experimentally, to confine light at nanometric scales. However, in practice, the nanopatterning of 2D materials often leads to a substantial degradation of its quality, reducing the associated polariton lifetime.

Here, we investigate 2D polaritons~\cite{Basov:2016,Low:2017,Basov:2021} sustained by parabolic waveguides based on 2D materials, which constitute a novel two-dimensional platform for strong light--matter interactions beyond conventional planar or layered structures. We theoretically demonstrate that the deformation of an otherwise flat 2D material into a parabolic shape can form an effective polaritonic waveguide, supporting modes which are laterally and normally confined, but propagate in the direction of translational symmetry (i.e., along the vertex of the channel). Our theoretical findings complement related proposals to achieve polariton waveguiding with 2D polariton-supporting media by engineering its supporting substrate, e.g., by depositing an atomically thin material on dielectric wedges or grooves~\cite{gonccalves2016graphene,Goncalves:2017,You:2019,Ye:2018}. In particular, the mechanical flexibility of an ultrathin film can be exploited to modify its plasmon resonances, as demonstrated in the microwave regime using flexible metal strips~\cite{Shen:2013} and predicted for crumpled graphene~\cite{Kang:2018}. Moreover, we show that the field confinement near the parabola vertex can promote large enhancements in the emission rate of an emitter, and thus can be readily probed in near-field microscopy~\cite{barcelos2015graphene,fei2016ultraconfined} or harnessed to control spontaneous emission of quantum light sources. We envision that our results can inspire and foster further experimental and theoretical efforts to realize both electrical and mechanical tuning of 2D channel polaritons in novel 2D-material-based nanophotonic architectures.

\section{Results and Discussion}

\begin{figure}[t!]
\includegraphics[width=0.5\textwidth]{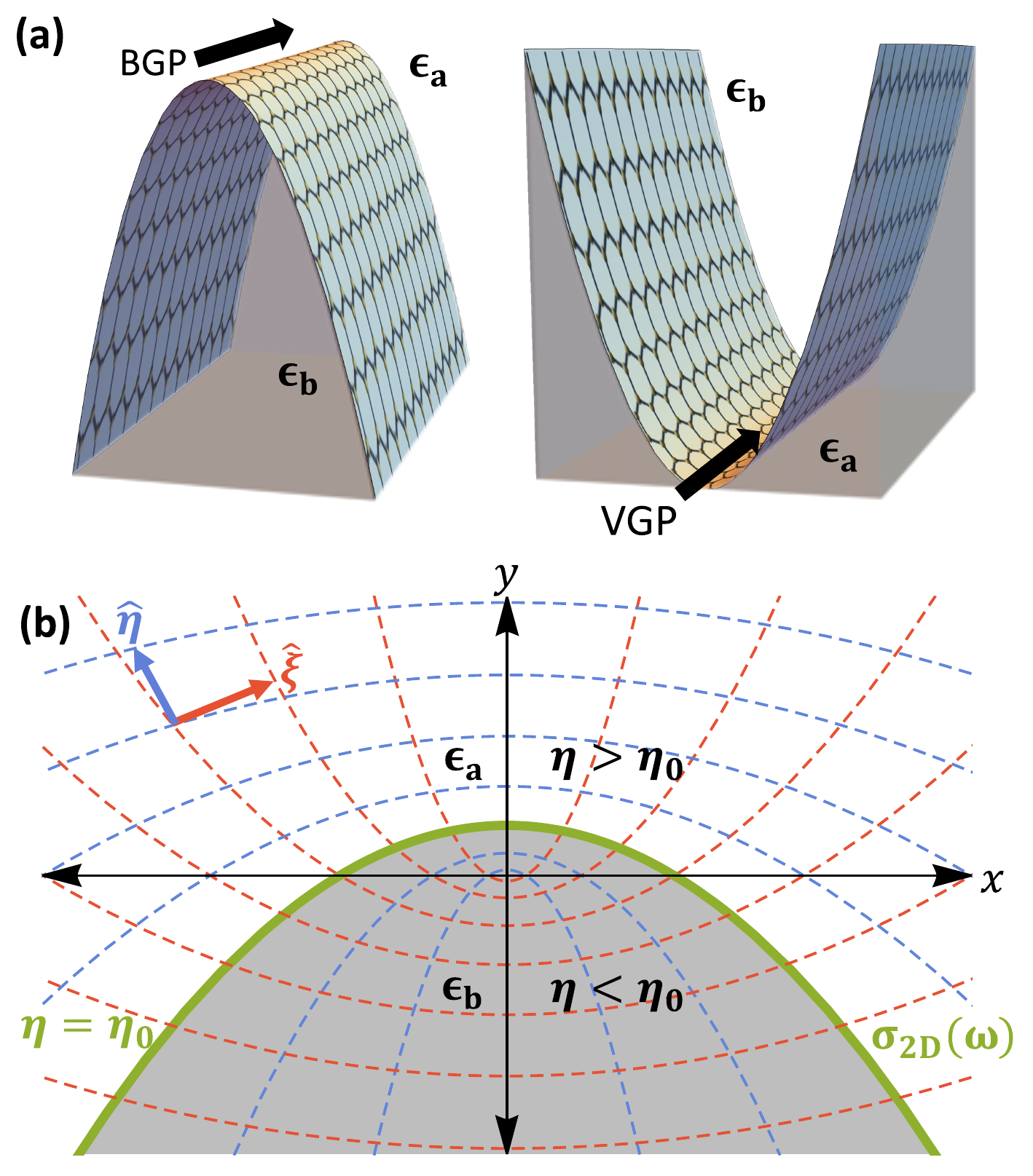}
    \caption{\textbf{Parabolic 2D-material-based waveguides.} \textbf{(a)} Schematic illustration of graphene (for concreteness purposes alone) parabolic waveguides, where the 2D material separates regions with relative permittivity $\epsa$ and $\epsb$, that can describe a bulge (or wedge) waveguide when e.g., $\epsa=1$ and $\epsb>1$ (left), or a valley (or groove) waveguide in the inverted situation (right). \textbf{(b)} Parabolic cylinder coordinate system superimposed on Cartesian coordinate system, where regions of dielectric permittivity $\epsa$ and $\epsb$ are bounded by $\eta > \eta_0$ and $\eta < \eta_0$, respectively, and the plane $\eta=\eta_0$ is characterized by a surface conductivity $\sigma_{\text{2D}}(\omega)\equiv\sigma(\omega)$.}
\label{fig:schematic}
\end{figure}

\noindent 
We consider a 2D material in the shape of a parabola, characterized by a two-dimensional (surface) conductivity $\sigma(\omega)$, that interfaces dielectric media above and below with relative permittivity $\epsa$ and $\epsb$, respectively [Fig.~\ref{fig:schematic}(a)]. Given the geometry under investigation, it is advantageous to work in a parabolic cylinder coordinate system $(\xi,\eta,z)$, which is related to the Cartesian coordinate system through~\cite{MorseFeshbach}
\begin{subequations}
\begin{align}
    x &= \xi \eta, \label{trans_1} \\
    y &= \frac{1}{2} \left( \eta^2 - \xi^2 \right) \label{trans_2}, \\
    z &= z,
\end{align}\label{eq:Cart_to_parabol}%
\end{subequations}%
where $- \infty < \xi < \infty$ and $0 \leq \eta < \infty$. In what follows, we study highly confined 2D polaritons, characterized by large wavevectors extending well-beyond the light-line, so that retardation effects become negligible and the electrostatic limit is well-justified. As we are chiefly concerned with electrostatic modes near the parabola's apex, we further assume that the parabola extends to infinity.

\noindent 
{\bf Polariton Dispersion.} Since the parabolic waveguide is translationally invariant along the $z$-direction, we may decompose the potential as ${\Phi(\xi,\eta,z) = \phi (\xi , \eta) \ee^{\ii q z}}$, where a harmonic time dependence of the form $\ee^{-\ii \omega t}$ shall be implicitly assumed henceforth. Denoting $\eta=\eta_0$ as the surface occupied by the 2D material and $\rho_{\text{2D}}(\xi,\eta=\eta_0)$ its associated 2D charge density, the system's polaritonic modes correspond to solutions of the Poisson equation
\begin{align}
    \left[ \frac{1}{\xi^2+\eta^2}\left(\frac{\partial^2 }{\partial \xi ^2} +  \frac{\partial^2 }{\partial \eta^2}\right) - q^2 \right] \phi (\xi, \eta) \nonumber\\
    = -\frac{\rho_{\text{2D}}(\xi, \eta_0 )}{\epsilon_0} \frac{\delta(\eta-\eta_0)}{\sqrt{\xi^2+\eta^2}} .
\end{align}
The solution to the above inhomogeneous partial differential equation (PDE) can be expressed as a linear combination of solutions of the corresponding homogeneous PDE~\cite{riley2006mathematical}, namely 
\begin{equation} \label{eq:phi}
    \phi(\xi,\eta) = \sum_n F_n(\xi) \begin{cases} A_n \, U\left(n+\tfrac{1}{2},\eta\sqrt{2q} \right), \, \eta>\eta_0 \\ B_n \, V\left(n+\tfrac{1}{2},\eta\sqrt{2q} \right), \, \eta<\eta_0 \end{cases}
\end{equation}
where $U$ and $V$ are parabolic cylinder functions~\cite{abramowitz1948handbook} and $F_n(\xi)$ are functions containing the Hermite polynomials $H_n(x)$,
\begin{equation}
    F_n(\xi) \equiv \frac{\left(q/\pi\right)^{1/4}}{\sqrt{2^nn!}}H_n\left(\xi\sqrt{q}\right)\ee^{-q\xi^2/2},
\end{equation}
forming a complete and orthonormal basis set~\cite{eguiluz1976electrostatic,boardman1981retarded,garcia1985excitation}. Defining the scalar potential above and below the 2D material as $\phi_{\rm a}(\xi,\eta) \equiv \phi(\xi,\eta>\eta_0)$ and $\phi_{\rm b}(\xi,\eta) \equiv \phi(\xi,\eta<\eta_0)$, respectively, it follows from the boundary conditions---enforcing the continuity of the electric potential at $\eta=\eta_0$, together with a jump $ \hat{{\bm \eta}} \cdot \left. \left[ \epsb {\bm \nabla} \phi_{\rm b}(\xi,\eta) - \epsa {\bm \nabla} \phi_{\rm a}(\xi,\eta) \right] \right|_{\eta=\eta_0} = \rho_{\text{2D}}(\xi,\eta_0)/\epsilon_0 $---that 
\begin{subequations}
\begin{equation} \label{eq:BC_potential}
    A_m U_m^0 = B_m V_m^0 \equiv C_m
\end{equation}
and
\begin{align} \label{eq:BC_displacement}
    &\epsa A_m U_m^{\prime 0} - \epsb B_m V_m^{\prime 0} = \frac{-\ii\sigma(\omega)}{\sqrt{2q}\epsilon_0\omega} \\
    &\times \int^{\infty}_{-\infty}d\xi \frac{F_m(\xi)}{\sqrt{\xi^2+\eta^2_0}} \left[\frac{\partial^2\phi}{\partial\xi^2} - q^2\left(\xi^2+\eta^2\right)\phi\right]_{\eta=\eta_0}, \nonumber
\end{align}\label{eq:BCs}%
\end{subequations}%
after projecting with $F_m$ and integrating over $-\infty < \xi < \infty$ , while also expressing the surface charge density $\rho_\text{{2D}}(\xi,\eta_0)$ in terms of the electrostatic potential by combining the 2D continuity equation with Ohm's law:  $\rho_{\text{2D}}= \ii \omega^{-1} \sigma(\omega) \left[\left(\xi^2+\eta^2\right)^{-1}\partial^2_\xi-q^2\right]\phi$. Additionally, in Eqs.~(\ref{eq:BC_potential})--(\ref{eq:BC_displacement}), we have introduced ${W_n \equiv W \left(n+\frac{1}{2} , \mu  \right)}$, ${W_n^0 \equiv W \left(n+\frac{1}{2} , \mu_0  \right)}$, and ${W_n^{\prime 0} \equiv \partial W_n/\partial \mu \big\vert_{\mu = \mu_0}}$, where ${W \equiv \{ U,V \}}$ and ${\mu \equiv \eta \sqrt{2q}}$, to simplify the notation. From Eqs.~\eqref{eq:BCs}, and using the fact that ${\partial^2_\xi F_n = q\left(q\xi^2-2n-1\right)F_n}$, we arrive at the following eigenvalue problem:
\begin{subequations}
\begin{equation} \label{eq:eigenvalue}
    \frac{\sqrt{2}}{q}\frac{\ii\epsilon_0\omega}{\sigma(\omega) } C_m = \sum_n M_{mn} C_n,
\end{equation}
where the matrix elements
\begin{equation} \label{eq:M_matrix}
    M_{mn} = -\frac{U_m^0V_m^0\left(2n+1+\beta^2\right)}{\epsa U_m^{\prime 0}V_m^0 - \epsb U_m^0V_m^{\prime 0}} \int^\infty_{-\infty} d\zeta\frac{f_m(\zeta)f_n(\zeta)}{\sqrt{\zeta^2+\beta^2}}
\end{equation}%
\label{eq:eigenvalue_and_M}%
\end{subequations}%
are expressed in terms of dimensionless quantities, namely, $\zeta = \xi \sqrt{q}$, $\beta = \eta_0 \sqrt{q}$, and $f_n \equiv q^{-1/4}F_n$; we note that these matrix elements admit an analytical solution.

Incidentally, the roots of the quantity ${\epsa U_m^{\prime 0}V_m^0 - \epsb U_m^0V_m^{\prime 0}}$ appearing in the denominator of Eq.~\eqref{eq:M_matrix} correspond to the dispersion of guided modes in a conventional parabolic wedge geometry \cite{eguiluz1976electrostatic,garcia1985excitation}, with the denominator reducing to the Wronskian ${U_m(z)V'_m(z)-U_m'(z)V_m(z)=\sqrt{2/\pi}}$ when $\epsa=\epsb$~\cite{abramowitz1948handbook}.

The spectrum of 2D polaritons in parabolic channels is thus determined by the eigenvalues of the matrix $\mathbf{M}$ [see Eqs.~(\ref{eq:eigenvalue_and_M})]: $\lambda_{n} \equiv \lambda_{n}(q)$. Hence, the dispersion relation of 2D channel
polaritons in parabolic waveguides follows from the implicit condition
\begin{equation}
 \lambda_{n}(q) = \frac{\sqrt{2}}{q} \frac{\ii \omega_n \epsilon_0}{\sigma(\omega_n)} . \label{eq:DispRel_general}
\end{equation}

Specifying, for concreteness, a graphene parabola and assuming that its optical response is described by a lossless Drude-like conductivity ${\sigma(\omega) = \ii e^2 \EF/(\hbar^2 \pi \omega)}$, we obtain the dispersion relation
\begin{equation} \label{eq:dispersion}
    \hbar \omega_n(q) = \Omega_{\text{flat}}(q) \sqrt{ \frac{\epsa + \epsb}{\sqrt{2}} \, \lambda_{n}(q) } ,
\end{equation}
where $\Omega_{\text{flat}}(q) = \sqrt{4 \alpha \EF \hbar c q/(\epsa + \epsb)} $ is the nonretarded dispersion for a flat, planar graphene sheet with Fermi energy $\EF$ sandwiched between two semi-infinite dielectric media characterized by relative permittivities $\epsa$ and $\epsb$, and where $\alpha \simeq 1/137$ is the fine-structure constant.
While Eq.~\eqref{eq:dispersion} only takes into account a single band within a Drude-type treatment of graphene's dynamical conductivity, the effect of interband transitions could be swiftly incorporated by employing the well-known expression for graphene's optical conductivity including both intra- and interband transitions (via the Kubo formula in the local-response approximation)~\cite{gonccalves2016graphene,Falkovsky:2008}, and solving Eq.~\eqref{eq:DispRel_general} numerically. Further, we assume here that the curvature of the parabola is small enough to neglect modifications in the conductivity introduced by strain or inhomogeneous doping (e.g., charge puddles) near the parabola vertex. More concretely, we restrict our investigation to parabolas of curvature $a\leq1$\,nm$^{-1}$, a threshold value inspired by explorations of plasmons in highly doped carbon nanotubes of $\gtrsim1$\,nm radius, for which the optical response is reasonably well-described by local conductivity of extended graphene~\cite{martin2015ultraefficient,devega2016plasmons}.

\begin{figure}[t!] 
\includegraphics[width=0.5\textwidth]{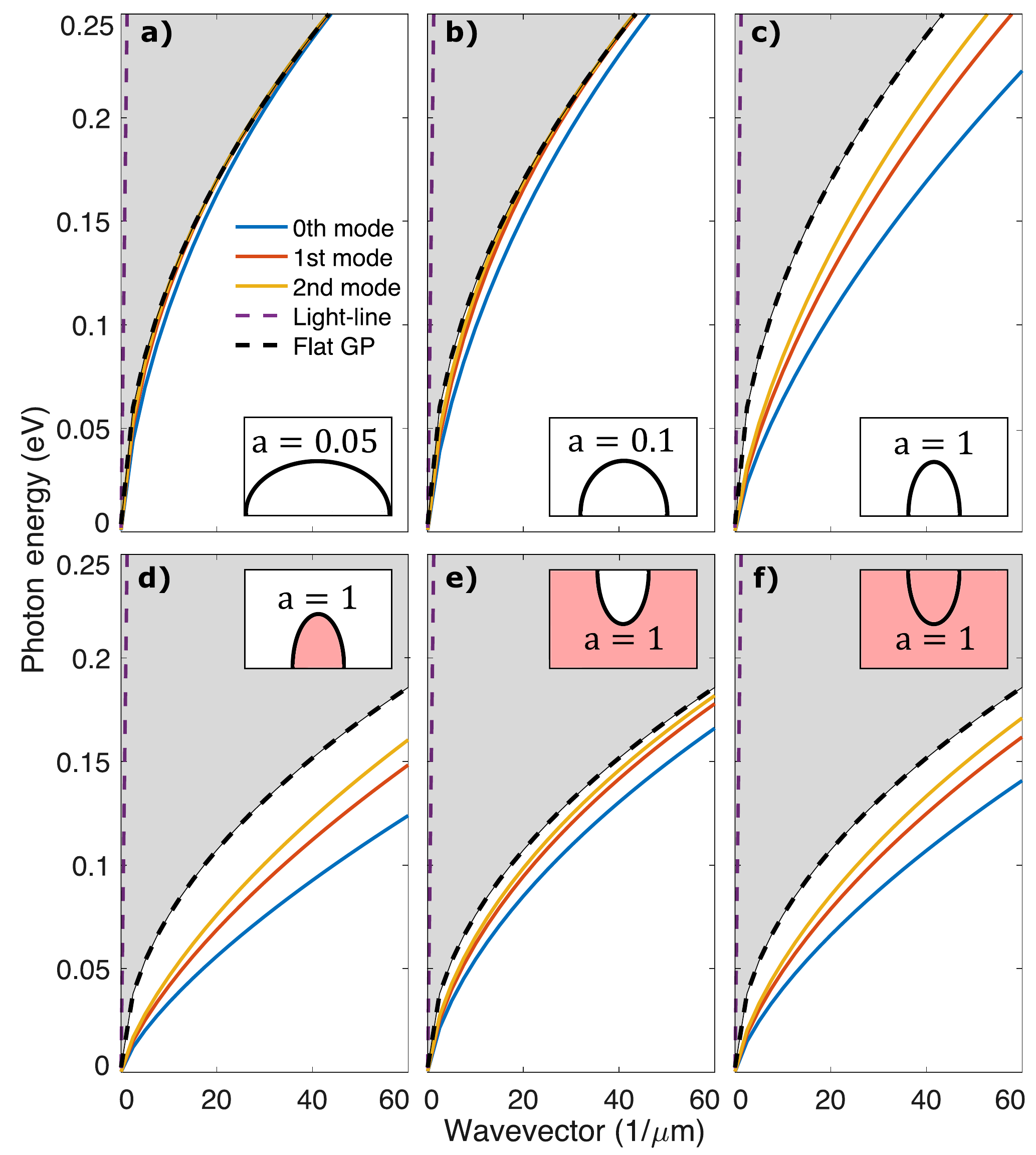}
\caption{\textbf{Plasmon dispersion in parabolic graphene waveguides.} \textbf{(a--c)} Plasmon dispersion [Eq.~\eqref{eq:dispersion}] for the three lowest-energy confined modes supported by a doped ($\EF=0.5$\,eV) graphene parabolas in vacuum ($\epsa=\epsb=1$) with the curvatures $a$ (in nm$^{-1}$) indicated in the insets.
\textbf{(d)}, \textbf{(e)}, and \textbf{(f)} show the dispersion relation for the parabolic bulge ($\epsa=1$, $\epsb=4$), parabolic valley ($\epsa=4$, $\epsb=1$), and homogeneous dielectric ($\epsa=\epsb=2.5$) configuration, respectively. 
In all panels, the graphene channel plasmon dispersion curves are color-coded in accordance with the legend in panel (a). The dispersion relation for flat graphene is indicated by a black dashed curve. The purple dashed line indicates the light-line in vacuum; the fact that it is nearly vertical and close to the vertical axis is direct consequence of the outstanding field confinement provided by graphene plasmons (thus justifying the quasi-static framework employed here).}
\label{fig:dispersion}
\end{figure}

We compute the eigenvalues $\lambda_{n}$ by solving the eigenvalue problem~\eqref{eq:eigenvalue_and_M} using standard linear algebra methods, and then obtain the ensuing 2D CPP dispersion relation via Eq.~\eqref{eq:dispersion}, which we present in Fig.~\ref{fig:dispersion} for the first three confined 2D CPP modes (in practice, convergence is achieved by truncating the expansion of Eq.\ \eqref{eq:phi} at $N\sim25$ terms). In Fig.~\ref{fig:dispersion}(a-c) we study a graphene parabola in vacuum ($\epsa=\epsb=1$) characterized by different curvatures $a\equiv 1/(2\eta_0^2)$ [notice that, from Eqs.~\eqref{eq:Cart_to_parabol}, one may write the parabola cross-section as $y = - ax^2 + \eta_0^2/2$]. The spectrum is comprised of a manifold of modes, which, for the same frequency, exhibit larger wavevectors $q$ and thus stronger field confinement with increasing curvature $a$. 
In Fig.~\ref{fig:dispersion}(d-f) we consider the effect of the surrounding dielectric environment for a fixed curvature (i.e., $a=1$\,nm$^{-1}$), and compare two distinct configurations: bulge graphene plasmons (BGPs), when $\epsa<\epsb$, and valley graphene plasmons (VGPs), characterized in our formalism by $\epsa>\epsb$. Of the cases considered, BGPs are found to exhibit the strongest confinement and VGPs the weakest, while all three configurations provide larger field confinements compared to the parabola in vacuum due to dielectric screening. Importantly, Fig.~\ref{fig:dispersion} shows that parabolic graphene waveguides support plasmonic modes with significantly larger wavevectors, and commensurably strong field confinement, than conventional plasmons in flat graphene. 

\begin{figure}[t!] 
\includegraphics[width=0.5\textwidth]{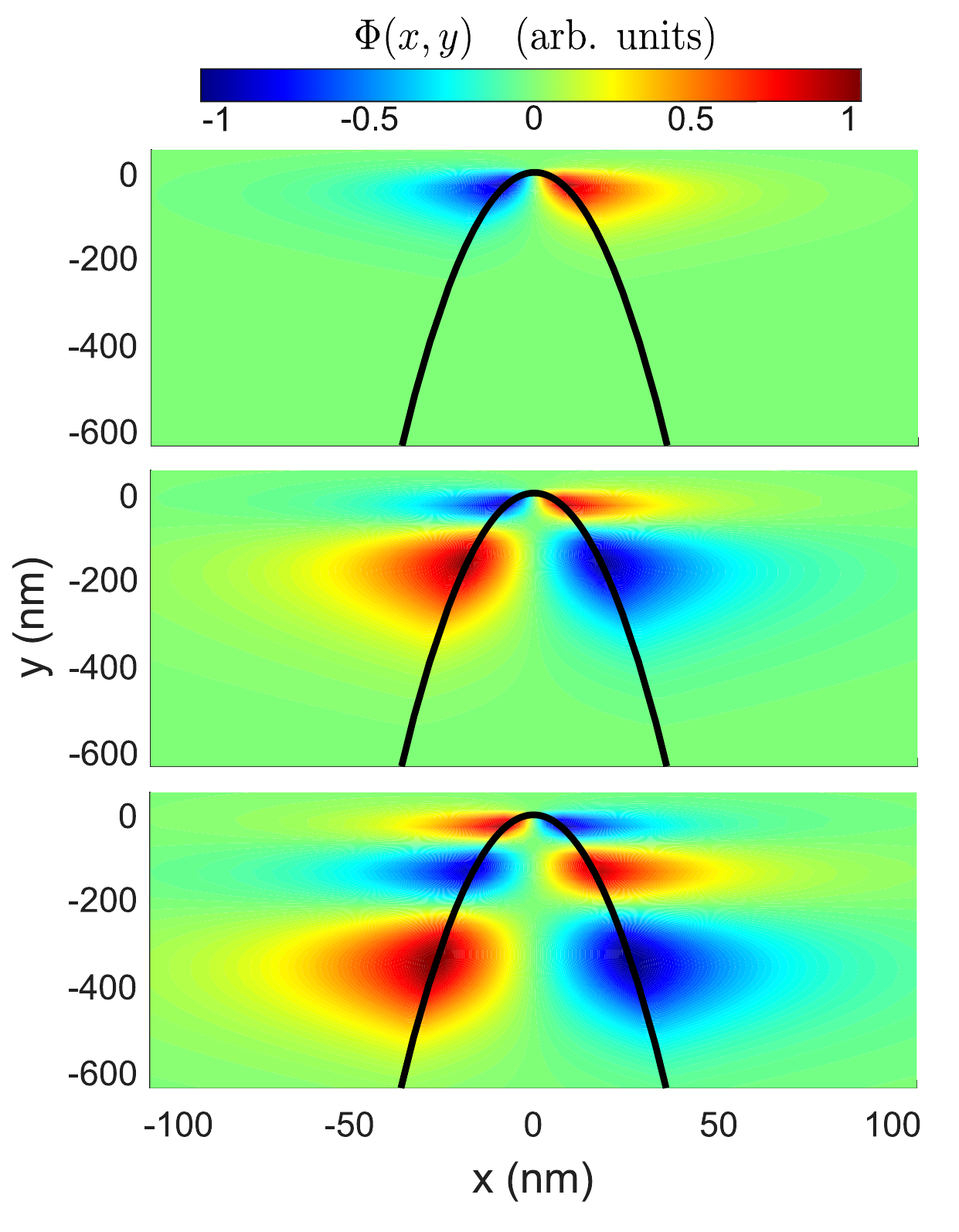}
\caption{\textbf{Electric potential associated with graphene channel plasmons in parabolic waveguides.} Cross-section of the electric potential, $\phi(x,y)$ [recall that the $z$-dependence is trivial, i.e., ${\Phi(\xi,\eta,z) = \phi (\xi , \eta) \ee^{\ii q z}}$], corresponding to the three lowest-energy 2D CPPs supported by a parabolic graphene waveguide of curvature $a=0.5$\,nm$^{-1}$, in vacuum ($\epsa=\epsb=1$) and doped to a Fermi energy $\EF=0.5$\,eV, for an excitation energy $\hbar\omega=0.2$\,eV and wavevectors $q=46\, \mu\text{m}^{-1}$, $37\, \mu\text{m}^{-1}$, and $34\, \mu\text{m}^{-1}$ in the upper, middle, and lower panels, respectively, where the potential in each panel is normalized to its maximum value.}
\label{fig:potential}
\end{figure}

Having solved the eigenvalue problem~\eqref{eq:eigenvalue_and_M}, the computed coefficients $A_n$ and $B_n$ (stemming from the eigenvectors of $\mathbf{M}$) can be inserted back into Eq.~\eqref{eq:phi} to construct the associated electric potential for the $n$-th mode. Figure~\ref{fig:potential} shows the cross-sections (in the $xy$-plane) of the potential associated with the three lowest-energy 2D CPP modes supported by a graphene parabolic channel with curvature $a=0.5\,\text{nm}^{-1}$. The lowest mode exhibits the largest field confinement and is very localized at the waveguide's vertex: the potential is mainly concentrated below $\sim100$\,nm distances from the graphene sheet, even though the wavelength of a photon with the same frequency in free-space is about $\lambda_0 \approx 6\,\mu\text{m}$, thereby underscoring the high field confinement that 2D CPP can achieve. For higher-order modes, the field confinement becomes progressively weaker, in agreement with the trend shown in Fig.~\ref{fig:dispersion}. 
Hence, the lowest-energy (fundamental) 2D channel plasmon mode attains the largest field confinement and thus is preferable when it comes to the squeezing of light below the diffraction limit; similar findings have been reported in theoretical studies of graphene-covered wedges and grooves~\cite{gonccalves2016graphene,Goncalves:2017} (and also on traditional metal-based plasmonic waveguides~\cite{moreno2006channel,Bozhevolnyi:2009,smith2015gap}). Finally, while we have assumed for simplicity that the parabola extends infinitely in the $y$-direction, we emphasize that our results can be used to characterize the spectral features of 2D CPPs in parabolic waveguides as long as the ``cut-off'' of the parabola is located well away from the tail of the electric potential.


\noindent {\bf Purcell Enhancement.} The intense electromagnetic field concentration provided by 2D polaritons in parabolic waveguides can be probed using scanning near-field optical microscopy (SNOM), as has been demonstrated for wrinkles and bubbles in graphene~\cite{barcelos2015graphene,fei2016ultraconfined}, or exploited for controlling the spontaneous emission rate $\Gamma$ of a nearby quantum emitter. In both cases, a relevant figure of merit is the Purcell factor~\cite{Novotny2006nanooptics},
\begin{equation} \label{eq:Purcell}
    \frac{\Gamma}{\Gamma_0} = 1+\frac{6\pi \epsilon_0 c^3 }{\vert\pb\vert^2\omega^3} \text{Im} \{ \pb^{*}\cdot\Eb^{\rm ind} \},
\end{equation}
where $\Gamma_0$ is the decay rate of the emitter in free space. Here, we compute the Purcell factor by considering the interaction of a parabolic graphene sheet occupying the $\eta=\eta_0$ plane with a point-dipole with moment $\pb$ located at $\rb'$, which generates an external potential
\begin{equation}
    \Phi = \frac{1}{4\pi\epsilon\epsilon_0}\pb\cdot\nabla'\frac{1}{\vert\rb-\rb'\vert}
\end{equation}
in a uniform medium with relative permittivity $\epsilon$. To describe the interaction, it is convenient to expand ${\vert\rb-\rb'\vert^{-1}}$ as (see Supporting Information)
\begin{align}
    &\frac{1}{\vert\rb-\rb'\vert} = \\
    &\pi^{1/2} \int^{\infty}_{-\infty} dq \frac{\ee^{\ii q(z-z')}}{\sqrt{q}} \sum_n F_n(\xi) F_n(\xi') U_n(\eta_>) V_n(\eta_<), \nonumber
\end{align}
where $\eta_>=\text{max}\{\eta,\eta'\}$ and $\eta_<=\text{min}\{\eta,\eta'\}$~\cite{garcia1985excitation}. Then, for a dipole emitter placed in the region above the parabola, the potential for a particular $q$-component is found by taking $\eta_>=\eta'$ and $\eta_<=\eta$, so that
\begin{subequations}
\begin{equation}
    \phi(\xi,\eta) = \sum_n F_n(\xi) 
    \begin{cases} A_n U_n(\eta) + \Delta_n V_n(\eta) , & \eta>\eta_0, \\ 
    B_n V_n(\eta), & \eta<\eta_0, \end{cases}%
\end{equation}%
where
\begin{equation}
    \Delta_n \equiv \frac{1}{2\epsa\epsilon_0}\sqrt{\frac{\pi}{q}} \pb\cdot\nabla'\left[F_n(\xi')U_n(\eta')\right] .
\end{equation}%
\end{subequations}%
Following the prescription leading to Eq.~\eqref{eq:eigenvalue}, we arrive at the matrix equation (see Supporting Information)
\begin{subequations}
\begin{align} 
    C_m + \frac{\epsa \sqrt{2/\pi} \Delta_m V_m^0}{\epsa U_m^{\prime 0} V_m^0 - \epsb U_m^0 V_m^{\prime 0}} \nonumber\\
    = -\frac{q}{\sqrt{2}}\frac{\ii\sigma(\omega)}{\epsilon_0\omega}\sum_n M_{mn} C_n,
     \label{eq:C_D_M_matrix}
\end{align}
where now
\begin{equation}
    C_m\equiv A_m U_m^0 + \Delta_mV_m^0 = B_mV_m^0.
\end{equation}
\end{subequations}
We solve Eq.\ \eqref{eq:C_D_M_matrix} through matrix inversion to compute the quantity
\begin{align}
    \pb^*\cdot\Eb^{\rm ind}&(\xi',\eta',0) =  \\
    &-\frac{2\epsa\epsilon_0}{\pi^{3/2}} \int^\infty_0 dq \sqrt{q} \sum_n \frac{\Delta_n}{U_n^0}\left(C_n - \Delta_n V_n^0\right)  \nonumber
\end{align}
entering Eq.~\eqref{eq:Purcell}, assuming without loss of generality that $z'=0$; a similar result can be obtained for a dipole positioned below the waveguide (see Supporting Information for details).

\begin{figure*}[ht!]
    \includegraphics[width=0.9\textwidth]{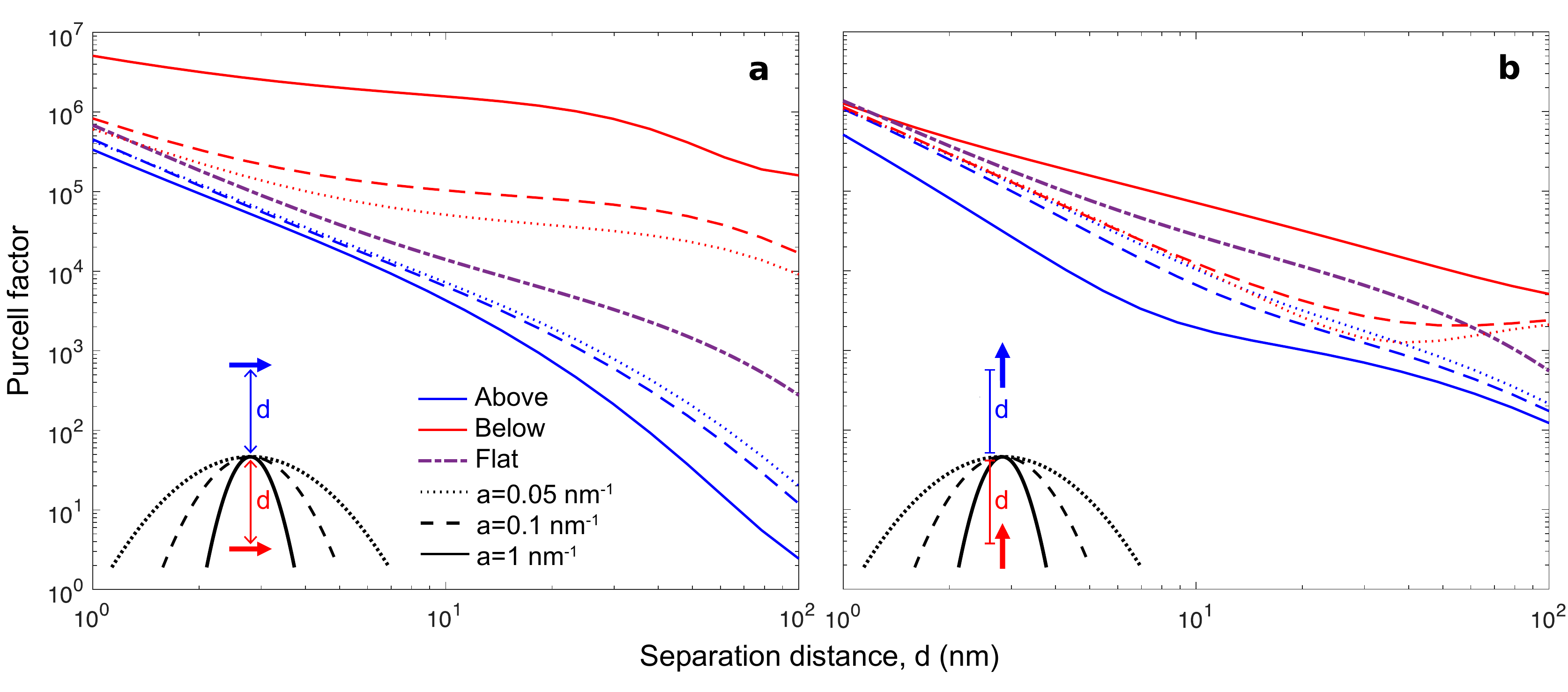}
    \caption{\textbf{Purcell enhancement.} Purcell factor experienced by a point-dipole emitting at $\hbar\omega=0.2$\,eV, positioned at a distance $d$ directly above (blue curves) or below (red curves) the parabola vertex, and oriented along $\hat{\bf x}$ (a) and $\hat{\bf y}$ (b). The results correspond to graphene parabolas with Fermi energy $\EF=1$\,eV and inelastic scattering rate $\hbar \gamma=0.05$\,eV in vacuum ($\epsa=\epsb=1$), characterized by the curvatures $a=0.05$\,nm$^{-1}$ (dotted curves), $0.1$\,nm$^{-1}$ (dashed curves) and $1$\,nm$^{-1}$ (solid curves).}
    \label{fig:Purcell_d}
\end{figure*}

We present in Fig.~\ref{fig:Purcell_d} the Purcell factor for dipole emitters oriented in either the $\hat{\bf x}$- (Fig.~\ref{fig:Purcell_d}a) or $\hat{\bf y}$- (Fig.~\ref{fig:Purcell_d}b) directions, and located at a distance $d$ above or below the vertex of the graphene parabolic-shaped waveguides (i.e., $\xi'=0$) with curvatures $a=0.05$\,nm$^{-1}$ (dotted line), $a=0.1$\,nm$^{-1}$ (dashed line) and $a=1$\,nm$^{-1}$ (solid line), along with the case of an emitter interacting with a flat graphene sheet to serve as a reference~\cite{Koppens:2011}. 

For an emitter aligned along $\hat{\bf x}$ (Fig.~\ref{fig:dispersion}a), our calculations predict a substantial enhancement of the Purcell factor with increasing curvature for an emitter below the waveguide, contrasting with the case where the emitter is placed above the parabolic waveguide, which exhibits the opposite behavior. Despite the stronger mode confinement of 2D CPPs when compared with their kins in planar graphene, we attribute this reduction in Purcell enhancement experienced by a dipole positioned above the parabola vertex to a diminishing of in-plane field component acting on the more ``folded'' graphene sheet, which is detrimental to the emitter--2D-CPP coupling. Conversely, positioning of the dipole directly below the parabola vertex results in stronger light--matter interactions, as the parabola, and its associated waveguide modes, inflect more rapidly towards the dipole as the curvature increases. In particular, the odd parity of the highly-confined waveguide modes facilitates stronger coupling of the induced field at the parabola vertex back on a dipole aligned along $\hat{\bf x}$. In contrast, the odd symmetry of these modes seems to be detrimental to the self-interaction of a dipole oriented in $\hat{\bf y}$, and typically leads to a reduction in the Purcell factor as indicated in Fig.~\ref{fig:Purcell_d}b; for extreme parabola curvature (e.g., $a=1$\,nm${^{-1}}$), bending of the 2D material facilitates coupling away from the vertex on each side of the dipole, providing some enhancement in the emission rate. Finally, we note that the Purcell factor for the emitter both above and below the waveguide approaches the value of a flat graphene sheet when $a \to 0$, as intuitively expected.

\section{Conclusions and Outlook}

We provide a semi-analytical formalism to describe 2D channel polaritons in one-dimensional waveguides with a parabolic cross-section. Here, we have placed special emphasis on graphene parabolas, but the present theoretical framework can be generally applied to different polariton flavors in other atomically thin media~\cite{Basov:2016,Low:2017,Basov:2021}, including ultrathin metal films~\cite{Chen:2016}. In the present analysis, we have neglected nonlocal effects in the graphene optical response, which is reasonable for a smoothly-deformed graphene sheet with a moderate radius of curvature. On the other hand, the possible singular response associated with an abrupt bending would potentially promote nonlocal effects in the response of graphene~\cite{Galiffi:2019}, similar to the situation in metallic grooves~\cite{Toscano:2013}.

Furthermore, the 2D parabola considered here constitutes the leading-order perturbation to an otherwise flat geometry, and thus can approximate the shape of a 2D material deposited onto wedges or valleys previously carved on a substrate, or suspended 2D materials in the shape of catenary curves which closely resemble parabolas. Infrared near-field spectroscopic methods, which have been successfully employed to probe the ultraconfined plasmon modes supported by wrinkles and nanobubbles forming in exfoliated graphene layers deposited on hexagonal boron nitride~\cite{barcelos2015graphene,fei2016ultraconfined}, could similarly be used to probe the spectral features of polaritons in curved suspended graphene. The combined ability to mechanically and electrically tune the long-lived and highly confined plasmonic excitations supported by graphene presents numerous possibilities for active engineering of nanoscale light--matter interactions.

\section*{Supporting Information}
This material is available free of charge via the internet at http://pubs.acs.org.

Expansion of the Green's function in parabolic cylindrical coordinates, an alternative approach to obtaining the self-consistent potential for an isolated parabola, and a detailed derivation of the Purcell factor for a dipole emitter near the parabola vertex.

\section*{Acknowledgments}
J.~D.~C. is a Sapere Aude research leader supported by Independent Research Fund Denmark (grant no. 0165-00051B).
S.~X. acknowledges the support from Independent Research Fund Denmark (project no. 9041-00333B).
S.~H. acknowledges support from the EU through ERC consolidator grant RYD-QNLO (Grant No. 771417) and the ErBeStA project (Grant No. 800942).
N.~A.~M. is a VILLUM Investigator supported by VILLUM FONDEN (grant no. 16498).
The Center for Nano Optics is financially supported by the University of Southern Denmark (SDU 2020 funding).
The Center for Nanostructured Graphene is sponsored by the Danish National Research Foundation (project no. DNRF103).
%


\providecommand{\latin}[1]{#1}
\makeatletter
\providecommand{\doi}
  {\begingroup\let\do\@makeother\dospecials
  \catcode`\{=1 \catcode`\}=2 \doi@aux}
\providecommand{\doi@aux}[1]{\endgroup\texttt{#1}}
\makeatother
\providecommand*\mcitethebibliography{\thebibliography}
\csname @ifundefined\endcsname{endmcitethebibliography}
  {\let\endmcitethebibliography\endthebibliography}{}

\end{document}